\documentstyle[11pt]{article}
\begin{document}
\addtolength{\baselineskip}{.45mm}
\input epsf
\newcommand{\vev}[1]{\langle #1 \rangle}
\def\mapright#1{\!\!\!\smash{
\mathop{\longrightarrow}\limits^{#1}}\!\!\!}
\newcommand{\bigoint}{\displaystyle \oint}
\newlength{\extraspace}
\setlength{\extraspace}{2mm}
\newlength{\extraspaces}
\setlength{\extraspaces}{2.5mm}
\newcounter{dummy}
\newcommand{\be}{\begin{equation}
\addtolength{\abovedisplayskip}{\extraspaces}
\addtolength{\belowdisplayskip}{\extraspaces}
\addtolength{\abovedisplayshortskip}{\extraspace}
\addtolength{\belowdisplayshortskip}{\extraspace}}
\newcommand{\ee}{\end{equation}}
\newcommand{\figuur}[3]{
\begin{figure}[t]\begin{center}
\leavevmode\hbox{\epsfxsize=#2 \epsffile{#1.eps}}\\[8mm]
\parbox{15cm}{\small\ 
\it #3}
\end{center} \end{figure}\hspace{-3.5mm}}
\newcommand{\fig}{{\it fig.}\ }
\newcommand{\newsection}[1]{
\vspace{15mm}
\pagebreak[3]
\addtocounter{section}{1}
\setcounter{subsection}{0}
\setcounter{footnote}{0}
\noindent
{\Large\bf \thesection. #1}
\nopagebreak
\medskip
\nopagebreak}
\newcommand{\newsubsection}[1]{
\vspace{1cm}
\pagebreak[3]
\addtocounter{subsection}{1}
\addcontentsline{toc}{subsection}{\protect
\numberline{\arabic{section}.\arabic{subsection}}{#1}}
\noindent{\large \bf \thesection.\thesubsection. #1}
\nopagebreak
\vspace{2mm}
\nopagebreak}
\newcommand{\ba}{\begin{eqnarray}
\addtolength{\abovedisplayskip}{\extraspaces}
\addtolength{\belowdisplayskip}{\extraspaces}
\addtolength{\abovedisplayshortskip}{\extraspace}
\addtolength{\belowdisplayshortskip}{\extraspace}}
\newcommand{\one}{{\bf 1}}
\newcommand{\zbar}{\overline{z}}
\newcommand{\ea}{\end{eqnarray}}
\newcommand{\is}{& \!\! = \!\! &}
\newcommand{\hf}{{1\over 2}}
\newcommand{\del}{\partial}
%
%
\newcommand{\twomatrix}[4]{{\left(\begin{array}{cc}#1 & #2\\
#3 & #4 \end{array}\right)}}
\newcommand{\twomatrixd}[4]{{\left(\begin{array}{cc}
\displaystyle #1 & \displaystyle #2\\be2mm]
\displaystyle  #3  & \displaystyle #4 \end{array}\right)}}
\newcommand{\low}{{\strut low}}
\newcommand{\hi}{{\strut hi}}
\newcommand{\ie}{{\it i.e.\ }}
\newcommand{\gbar}{{\overline{g}}}
\newcommand{\half}{{\textstyle{1\over 2}}}
\newcommand{\tfrac}{\frac}
\renewcommand{\thesubsection}{\arabic{subsection}}
\begin{titlepage}
\begin{center}

{\hbox to\hsize{
\hfill PUPT-1911}}

\bigskip

\bigskip

\bigskip

\bigskip

\vspace{6\baselineskip}

{\Large \sc On Open/Closed String Duality}

\bigskip

\large{

\bigskip
\bigskip
{\sc  Justin Khoury\footnote{jkhoury@princeton.edu} and Herman 
Verlinde\footnote{verlinde@feynman.princeton.edu}}\\[1.1cm]

{ \it Joseph Henry Laboratories, Princeton University,
Princeton, NJ 08544}\\[1.4cm]

\vspace*{1.5cm}

{\bf Abstract}\\}

\end{center}

\noindent 
It was recently shown, using the AdS/CFT correspondence, that the low energy
effective action of a large $N$ open string theory satisfies a holographic RG
flow equation closely related to the Hamilton-Jacobi equation of 5-d
supergravity.  In this paper we re-obtain the same flow equation in the dual
regime of small 't Hooft coupling $\lambda\ll 1$.  Our derivation makes use of
the conformal equivalence between planar open string diagrams and closed string
tree diagrams.  This equivalence can be viewed as a microscopic explanation of
the open/closed string duality that underlies the AdS/CFT correspondence.

\end{titlepage}

\noindent{\bf 1. Introduction}

\smallskip

In this paper we will consider the world volume theory of a set of $N$ D3
branes, in the limit of large $N$, at {\it finite} $\alpha'$. 
When embedded in flat 10-d space-time, the
theory  reduces at low energies to ${\cal N}=4$ supersymmetric Yang-Mills theory. 
More generally, we can deform the model by turning on gauge invariant couplings
$\phi^{i}$ and consider the quantum partition function as a function of the
$\phi^{i}$
\be
\exp\Bigl({i\, \Gamma(\phi)}\Bigr)\,=\,\Bigl\langle\,\exp\Bigl({i\! \int\! \phi^{i}O_{i}}\,\Bigl) \Bigr\rangle.
\ee
Via the famous AdS/CFT correspondence \cite{adscft}, this partition function 
has, for $\alpha' \to 0$ and for large 't Hooft coupling 
$\lambda =Ng_{ym}^{2}$, a 
dual represention as that of
IIB string theory on (a deformation of) $AdS_{5}\times\,S^{5}$. In this
correspondence, the couplings $\phi^{i}$\ \ -- \ which among others include
the dilaton, 4-d space-time metric, as well as RR fields -- \ specify the non-dynamical
asymptotic values of the corresponding set of closed string fields \cite{gkp} \cite{ew}. When the
$\phi^i$ represent finite expectation values, they generally break the conformal
invariance and part or all of the supersymmetry of the low energy theory.

An interesting generalization of this duality arises when all directions
perpendicular to the D3-branes are taken to be compact \cite{hv}. In this case one can
not take the decoupling limit and interactions between open and closed strings
will remain relevant. Instead, the target space of the IIB string theory is
described by a warped compactification similar to the Randall-Sundrum
geometry \cite{rs}, and variations of the couplings $\phi^{i}$\ correspond to
normalizable, and thus dynamical, fluctuations of closed string modes. The low energy dynamics 
of these modes is described by an effective action $S(\phi)$, that includes besides the
non-local induced action $\Gamma(\phi)$\ also a local contribution $S_{E}(\phi)$ arising
from the KK reduction of the 10-d effective action of the IIB string
theory
\be
S(\phi)\,=S_{E}(\phi\,)+\,\Gamma(\phi).
\ee
In \cite{vv} some properties of this effective action where studied for large 
$\lambda$ with the
help of the AdS/CFT correspondence. Using the dual supergravity approximation
and elementary results of Hamilton-Jacobi theory, it was found that in this
regime $S(\phi)$ satisfies an RG flow equation of the following schematic form
\be
\label{bflo}
\beta^{i}(\phi)\frac{\partial S}{\partial\phi^{i}}-\frac{1}{2}\,G^{ij}
\frac{\partial S}{\partial\phi^{i}}\,\frac{\partial S}{\partial\phi^{j}%
}\,=\,0.
\ee
Here $G^{ij}$ denotes some apropriate metric on the space of couplings, and $\beta^{i}(\phi)$ 
are `beta-functions' that describe the classical flow velocities
of the $\phi$-fields as a function of the holographic extra dimension. They
are expressed in terms of the local action $S_{E}(\phi)$ as
\be
\beta^{i}(\phi)\,={G}^{ij}\frac{\partial S_{E}}{\partial\phi^{i}}.\,
\ee 
For a more detailed explanation of these relations we refer to \cite{vv}.

In \cite{vv} eqn (3) was interpreted as an RG invariance of the total gravitational 
effective action $S$, in the sense that any classical extremum of $S$
automatically lies on a complete RG trajectory of classical solutions connected by the flow
relation
\be
\dot{\phi}^i = \beta^i(\phi).
\ee
This property of $S$, if true in general, could be of particular importance for the
cosmological constant problem.
The derivation of (3) as given in \cite{vv}, however, is valid only in the limit 
of large $N$ and large 't Hooft coupling $\lambda\gg 1$. It is important therefore
to investigate whether these relations can be generalized and extended to other regimes
as well. In the following we will concentrate on the dual regime of small $\lambda,$ but still 
infinite $N.$ We will find the positive result that the exact same equations  (2), (3) and (4) 
remain valid also in this regime, except that the quantities $\Gamma$, $\ S_{E}$, $G^{ij}$ 
and $\beta^{i}$ are now all defined via their weak coupling descriptions. 

The main idea behind our derivation is that the world sheet of a planar
multi-loop diagram in open string theory is conformally equivalent to a closed
string tree diagram. Indeed, all holes in the open string diagram can be
represented in the dual channel by means of external closed string states,
equal to the appropriate D-brane boundary state $|\,B\,\rangle$. Via this dual
representation all potential UV divergences of the open string diagram become
equivalent to potential IR divergences due to on-shell closed string states in
the dual channel. It is not too surprising therefore that the RG
structure of the large $N$  open string theory can be made to look
identical to a classical evolution equation of closed string theory.

Several elements in our reasoning have appeared in earlier works. We mention here:

\begin{itemize}

\vspace{-2.5mm}

\item  The Fischler-Susskind mechanism for cancelling string loop
divergences \cite{fs}; for a recent discussion of the FS-mechanism in relation
with D3-brane physics, see \cite{bl}.

\vspace{-2mm}

\item  The interpretation given in \cite{ram} of the BV symmetry of closed
string field theory \cite{csft} as an `exact' RG invariance \`a la
Polchinski \cite{p83}.

\vspace{-2mm}

\item  The non-linear flow relation,  proposed by Polyakov in \cite{p93}, satisfied by
the tree level partition function in a general non-critical string theory.
\end{itemize}

\vspace{-2mm}

\noindent 
In a pure field theory context, the structure described below also seems closely related 
to the old recursive formula for QFT counterterms due to Bogolyubov \cite{bogol}, that forms the basis for
the classic BPHZ renormalization method. We will comment on this correspondence (which may prove
useful for taking the $\alpha' \to 0$ limit of our result) in section 3.  
In section 4 we summarize our conclusions.

\bigskip\bigskip\bigskip

\noindent{\bf 2. Derivation of the flow equation}

\medskip

\noindent
For small 't Hooft coupling $\lambda$, the total low energy effective action
$S(\phi)$ is obtained by summing over all n-loop planar open string diagrams
in the closed string background specified by $\phi.$ Schematically\footnote{Here we 
are using the same notation $O_i$ for the closed string vertex operators dual to the
fields $\phi^i$, as used in eqn (1) for the gauge theory operators dual to $\phi^i$. In principle,
one can make a one-to-one correspondence between the two, by comparing the 2-d world-sheet 
action with the D3-brane world volume action in a given closed string background.  }
\be
\label{exp}
S(\phi)\,=\,\Gamma_{0}(\phi)\,+\,\sum_{\rm n\geq1}\,\lambda^{n}\,\Gamma_{\rm n}(\phi)
\ee%
\be
\,\Gamma_{\rm n}(\phi)\,=\Bigl\langle\,\exp\Bigl({i\! \int\! \phi^{i}O_{i}}\Bigr)
\,\Bigr\rangle\,_{\rm n}%
\ee
This term $\Gamma_{n}(\phi)$ is the ${\rm n\! -\! 1}$-loop open string contribution,
given by the partition function of the world-sheet sigma-model (parametrized 
by $\phi$) on a sphere with n holes, integrated over all moduli parameterizing 
the relative sizes and locations of these holes. Both the sigma-model expectation 
value and the
integral over these moduli may produce potentially infinite answers, which
both can be regularized by introducing an explicit cut-off scale $\epsilon$. 
We will give a concrete prescription for this cut-off momentarily. In any case,  
in the end all physical answers should, when expressed in terms of renormalized 
couplings, be independent of this cut-off. In particular, writing
\be
S(\phi(\epsilon);\epsilon)\,,
\ee
where $\phi(\epsilon)$ is the renormalized sigma-model background satisfying
the RG equation
\be
\label{rgf}
\epsilon\frac{\partial\phi^{i}}{\partial\epsilon}\,=\,\beta^{i}(\phi),
\ee
with $\beta^{i}(\phi)$ the sigma model beta-functions, we must require that
the total $\epsilon$ dependence cancels
\be
\label{canc}
\epsilon\frac{dS}{d\epsilon}\, =\, \beta^{i}(\phi)\frac{\partial S}{\partial\phi
^{i}}\,+\,\epsilon\frac{\partial S}{\partial\epsilon}\,=\,0.
\ee
This requirement reduces to the usual condition of conformal invariance in the
limit $\lambda\rightarrow 0$, when the holes of the open string loops are absent.
The condition (\ref{canc}) with $\lambda>0$ relates divergences arising from
deviation from world-sheet conformal invariance to divergences coming from the
shrinking of open string loops; the cancellation of these two different types of
divergences is known as the Fischler-Susskind mechanism \cite{fs}.

To implement this cancellation mechanism, we need a sufficiently precise
definition of the cut-off $\epsilon$, both for the divergences of the
sigma-model as well as for regulating the moduli integral.  For this we will
make use of some technology from closed string field theory \cite{csft}.  To any
Riemann surface (with holes) with given conformal structure, we can assign a
unique minimal area metric $g_{\alpha\beta}.$ For a given point on the moduli
space of the open string loop diagram, we can thus measure the minimal geodesic
length $\ell(C)$, as defined using this minimal area metric, of all
non-contractible contours $C$ surrounding a non-zero number of holes.  (See fig
1a).  The UV divergences of the loop diagram arise when one or more of these
geodesic lengths $\ell(C)$ tends to zero.  We will therefore introduce a UV
regulator $\epsilon$ by requiring that the moduli integral is restricted to
those conformal structures for which
\be
\label{rest}
\ell(C) \geq \epsilon
\ee
for all non-contractible contours $C.$ Hence the boundary of the regulated
moduli space are degenerate surfaces for which the above bound (\ref{rest}) is
saturated for one or more contours $C$.

\vspace{-4mm}

\figuur{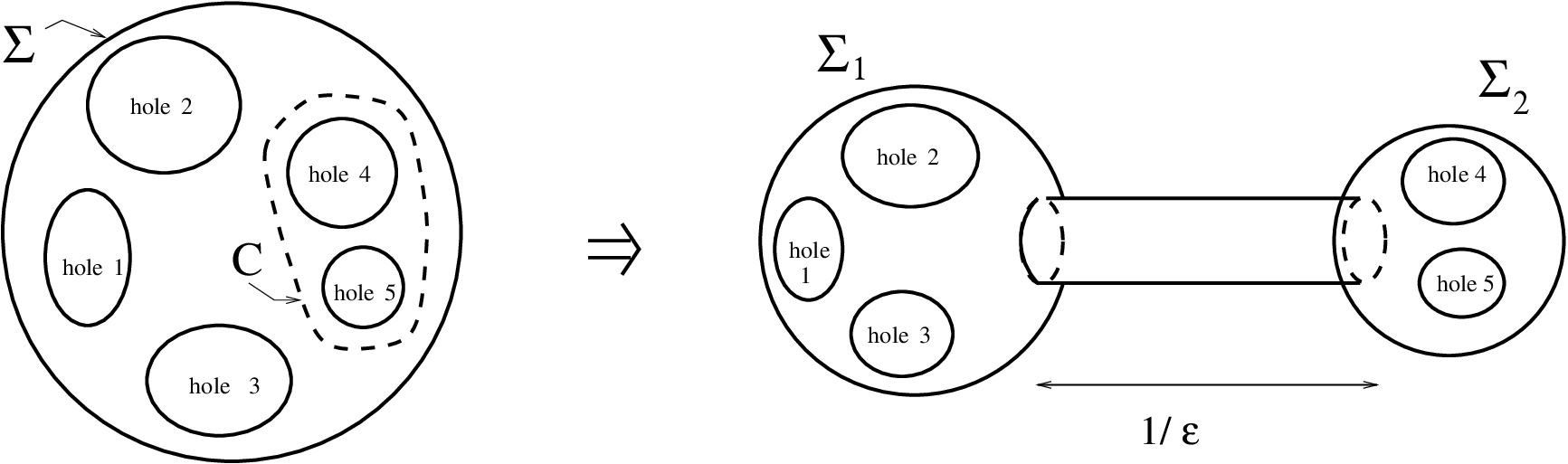}{12cm}{{\bf Fig 1a and 1b}: A planar n loop open string diagram is given by the integral over all 
shapes of a spherical Riemann surface $\Sigma$ with n+1 holes. When a non-contractible contour $C$
surrounding some of the holes acquires a very small length $\ell(C) =\epsilon$, the surface
degenerates into two separate spherical surfaces $\Sigma_1$ and $\Sigma_2$ connected by a
long tube of length $1/\epsilon$.}

Since in the end we need to compare this type of degeneration of the open string loop diagram
with the sigma-model divergences, it seems most practical to
regulate the sigma-model expectation values in an analogous fashion. To this
end, we explicitly expand the exponential in eqn (7)
\be
\label{ggn}
\Gamma_{\rm n}(\phi;\epsilon)=\sum_{k\geq0}\frac{1}{k!}\,\langle\,
\mbox{\raisebox{-8pt}{${{\underbrace{\Phi\cdots\Phi}}\atop{k \times}}$}}
\,\rangle_{\rm n}=\sum_{k\geq0}\frac{1}%
{k!}\,\langle\,(\Phi)^{k}\rangle_{\rm n}%
\ee
with
\be
\Phi=\sum_{i}\,\int\phi^{i}\,O_{i}\,.
\ee
The $k$-th order term on the right-hand side is a correlator, defined in the $\phi=0$ 
sigma-model, of \ $k$ operators $\Phi$ on an ${\rm n}\! - \! 1$ loop open string diagram. 
The resulting amplitude is therefore an integral over the moduli space of a sphere with $n$ holes and $k$
punctures. (See fig 2a.) We can now apply the same construction as above, and use the unique
minimal area metric on this punctured surface to assign a given minimal geodesic length
to all closed contours surrounding a non-zero number of holes and/or punctures,
and require that all such lengths must be larger than the cut-off
$\epsilon.$ In this way we have indeed introduced one uniform cut-off procedure 
for both types of divergences.\footnote{Given the limited available tools for
dealing with sigma-models with RR backgrounds, the procedure outlined here seems at 
present the only precise method for extracting the cut-off dependence of the sigma-model
expectation values. We should further mention that, in case the string theory under consideration
is an orientifold compactification, one also needs to include world-sheets with an arbitrary number of
cross-caps. These can be treated in a similar fashion as the open string holes and $\Phi$ 
operator insertions.}

The above equation (\ref{ggn}) in fact needs some extra specification for the
case $n=0$. On the sphere without holes one needs, due to the invariance under the global conformal group
$SL(2,C)$, at least three operator insertions $\Phi$ to obtain a well-defined expectation value.
To write the $n=0$ contribution to the effective action, we must thus include a separate kinetic
term via
\be
\label{gzero}
\Gamma_0(\phi;\epsilon) = {1\over 2} \langle \Phi | Q |\Phi\rangle + 
\sum_{k\geq 3}\frac{1}%
{k!}\,\langle\,(\Phi)^{k}\rangle.
\ee
Here $Q$ denotes the nil-potent world-sheet BRST charge of the $\phi=0$
sigma-model and $|\Phi\rangle = \sum_i \phi^i | O_i\rangle$ is the state
corresponding to the sigma-model background $\Phi$.  The above expression for
$\Gamma_0(\phi)$ is of the same form as the standard classical action of closed
string field theory (for a detailed discussion of its definition and properties,
see \cite{csft}).  Note, however, that in the present context the classical
equations of motion of $\phi$ must be derived from the total action $S(\phi)$
given in (6), and not from just $\Gamma_0(\phi)$.  Indeed, since $\Gamma_0(\phi)$
has no contribution from surfaces with holes, it does not have any direct
knowledge of the presence of the D3 branes.

\vspace{-4mm}

\figuur{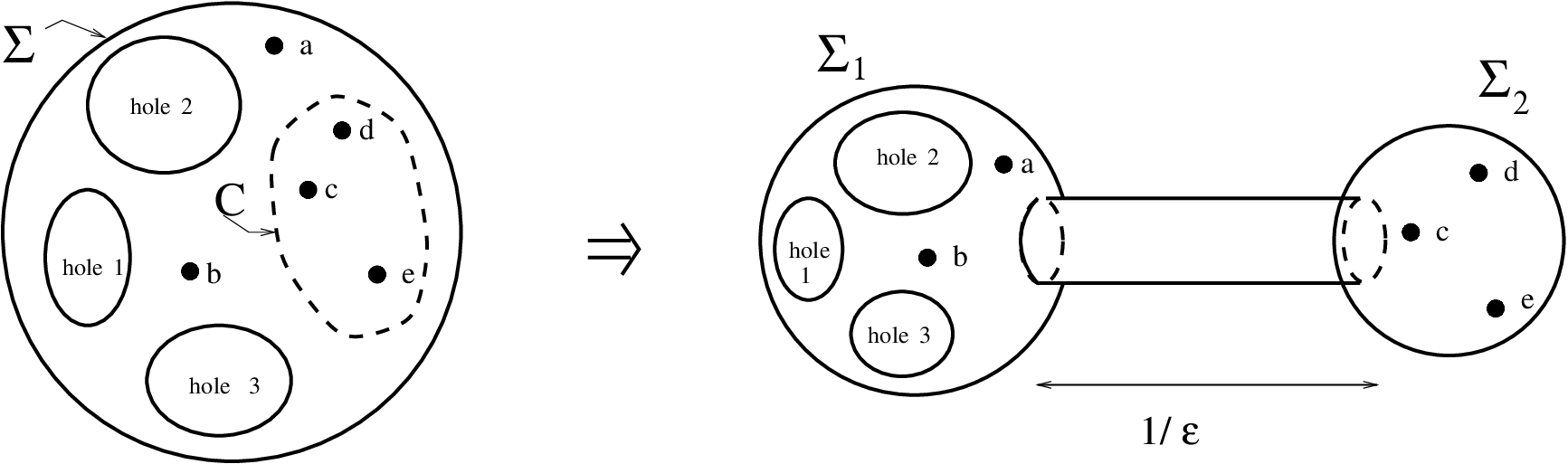}{12cm}{{\bf Fig 2a and 2b} After expanding the sigma-model interaction term
$\exp \Phi$, planar open string 
diagrams have in addition to holes also a number of insertions of the closed string operator $\Phi$ located
at isolated punctures. The sigma model divergences correspond 
to the degeneration of non-contractible curves $C$ that separate off spheres 
with no holes but with a non-zero number of punctures.}

It is now straightforward to determine the $\epsilon$ dependence of the total
effective action.  Each term $\langle (\Phi)^k \rangle_{\rm n}$ is given by an
integral over the corresponding moduli space, whose only dependence on $\epsilon$
is via the restriction (\ref{rest}) on the geodesic lengths.  Hence if we
differentiate $\langle (\Phi)^k \rangle_{\rm n}$ with respect to $\epsilon$, the
result is an integral over the boundary sub-space for which (at least) one of the
contours $C$ has reached its minimal length $\ell(C)=\epsilon$.  Now it's a well
known fact that such a pinched surface is conformally equivalent to a surface for
which the closed string propagator in the dual channel has acquired a large
length proportional to $1/\epsilon.$ (See figs 1b and 2b.)  The partition
function for this degenerate surface factorizes into a sum of products of two
one-point functions defined on each half of the surface on each side of this long
propagator.  Concretely, writing the evolution operator along this long tube as
\be
\label{prop}
\epsilon^{L_{0}+\overline{L}_{0}}= {1\over 2} \sum_{i}\,|O_{i}\rangle\,G^{ij}\,\langle O_{j}|,
\ee
(here we assume that the set of states $|O_i\rangle$ forms a complete basis of
closed string states) we can express the explicit $\epsilon$-dependence of the
${\rm n}$ loop partition function $\Gamma_{\rm n}$ as follows
\be
\label{gsum}
\epsilon\,\frac{\partial\Gamma_{\rm n}}{\partial\epsilon}\,= - {\rm{\,}\frac{1}
{2}\sum_{0\leq m\leq n}}\,G^{ij}\,\frac{\partial\Gamma_{\rm m}}{\partial\phi^{i}
}\,\frac{\partial\Gamma_{\rm n-m}}{\partial\phi^{j}}
\ee
The summation here runs over all possible ways of dividing the surface with 
${\rm n}$ holes into two parts, as indicated in figs 1b and 2b.

The ${\rm m} \! = \! 0$ and ${\rm m \! = n}$ terms in eqn (\ref{gsum}), the ones
containing a factor $\partial\Gamma_0/\partial\phi^i$, describe the `splitting 
off' of a sphere without holes as indicated fig 2b, and represent 
the scale dependence
due to the sigma model divergences.\footnote{Note that (via the presence of the
kinetic term in (\ref{gzero})) $\partial\Gamma_0/\partial\phi^i$ starts out for
small $\phi$ with a linear term proportional to $\Delta^{i}_{j} \phi^j$, with
$\Delta_j^i$ the matrix of scaling dimensions of the $O_i$ (defined via $L_0
|O_i\rangle = \Delta_{i}^j |O_j\rangle$).  The corresponding explicit cut-off
dependence of the correlators $\langle (\Phi)^k \rangle_{\rm n}$ arises from the
local regularization of the individual $\Phi$ operator insertions.}  Our method
of renormalization of the sigma model is to absorb this particular type of
divergence by means of the transition to renormalized couplings
$\phi^i(\epsilon)$.  In other words, the $\epsilon$ dependence of
$\phi^i(\epsilon)$ should be such that the terms proportional to
$\partial\Gamma_0/\partial\phi^i$ cancel in the total $\epsilon$-variation of
$\Gamma_n$.  From this we deduce that the renormalized couplings must satisfy
(\ref{rgf}) with
\be
\label{bg}
\beta^{i}(\phi)=G^{ij}\,\frac{\partial\Gamma_{0}}{\partial\phi^{j}}.
\ee
Note that a relation of this form is indeed expected from the identification
(\ref{gzero}) of $\Gamma_0(\phi)$ with the classical closed string field theory
action.

[ This renormalization procedure (\ref{bg}) is further motivated by the
requirement that the eventual renormalized form of the RG-equation should be
independent of the cut-off $\epsilon$.  In particular the factorization metric
$G^{ij}$ introduced in (\ref{prop}) should be cut-off independent.  It seems 
reasonable to assume that the only divergences, that induce a renormalization of 
$G^{ij}$, are those of the
sigma-model.  In our regularization method, these arise from `contact term'
contributions produced when one or more of the $\Phi$-operator insertions
get `trapped' somewhere on the long tube in fig 2b.  The effect of these terms is
to replace the $L_0$-operators in (\ref{prop}) by those in the $\Phi$-background.
The resulting condition of $\epsilon$-independence of $G^{ij}$ takes the form
\be
\epsilon \, {\partial G^{ij}\over \partial \epsilon}
+ \beta^k \, {\partial G^{ij}\over \partial \phi^k} - \Delta_k^i \,
G^{kj} - \, \Delta_k^j
G^{ik} = 0.
\ee
with $\beta^i$ as in (\ref{bg}) and where
$\Delta_i^k = \partial_i \beta^k$ denotes the matrix of conformal
dimensions of the $O_i$ in the $\Phi$-background. ]

Via the definition of the total action $S$ as the sum of all $\Gamma_n$'s, we 
can summarize the infinite set of relations (\ref{gsum}) as a single non-linear
equation for $S$.  
\be 
\epsilon \,\frac{\partial S}{\partial\epsilon}\,=\, - {1\over 2}
G^{ij}\frac{\partial S}{\partial \phi^{i}}\,\frac{\partial S}{\partial\phi^{j}},
\ee and we thus indeed obtain the announced form for the total FS scale
invariance condition (\ref{canc}) 
\be 
\label{final} \beta^{i}(\phi)\frac{\partial
S}{\partial\phi^{i}}\,- {1\over 2} G^{ij}\frac{\partial
S}{\partial\phi^{i}}\,\frac{\partial S}{\partial\phi^{j}}\,=\,0 .  
\ee 
This equation, as well as our derivation, are strongly reminiscent of
Polchinski's version of the Wilsonian renormalization group \cite{p83}.  
Indeed our UV-regulator (\ref{rest}) looks just like an IR cut-off on the (proper length 
of the) closed string propagator in the dual channel. It is not surprising
therefore that the explicit $\epsilon$-dependence of $S$ looks just like
(the classical limit of) the RG-equation derived in \cite{p93}.

Finally, we remark that the total equation of motion for $S$ implies that 
on-shell
\be
\label{this}
G^{ij} {\partial S \over \partial \phi^j} \, = \,
\beta^i(\phi) +  \sum_{{\rm n} \geq 1} \lambda^{\rm n}\; G^{ij} 
{\partial \Gamma_{\rm n} \over \partial \phi^j} = 0,
\ee 
which tells us that the deviation from scale-invariance coming from the
sigma-model beta-functions cancels against that coming from the source 
terms due to the presence of the D3-branes, and/or from the open string 
loop contributions.\footnote{The term with n=1 is a pure D3-brane source term; 
the other terms with n$>$1 can either be viewed as higher order closed string
source terms or as quantum corrections due to open string loops.}

\bigskip
\bigskip


\figuur{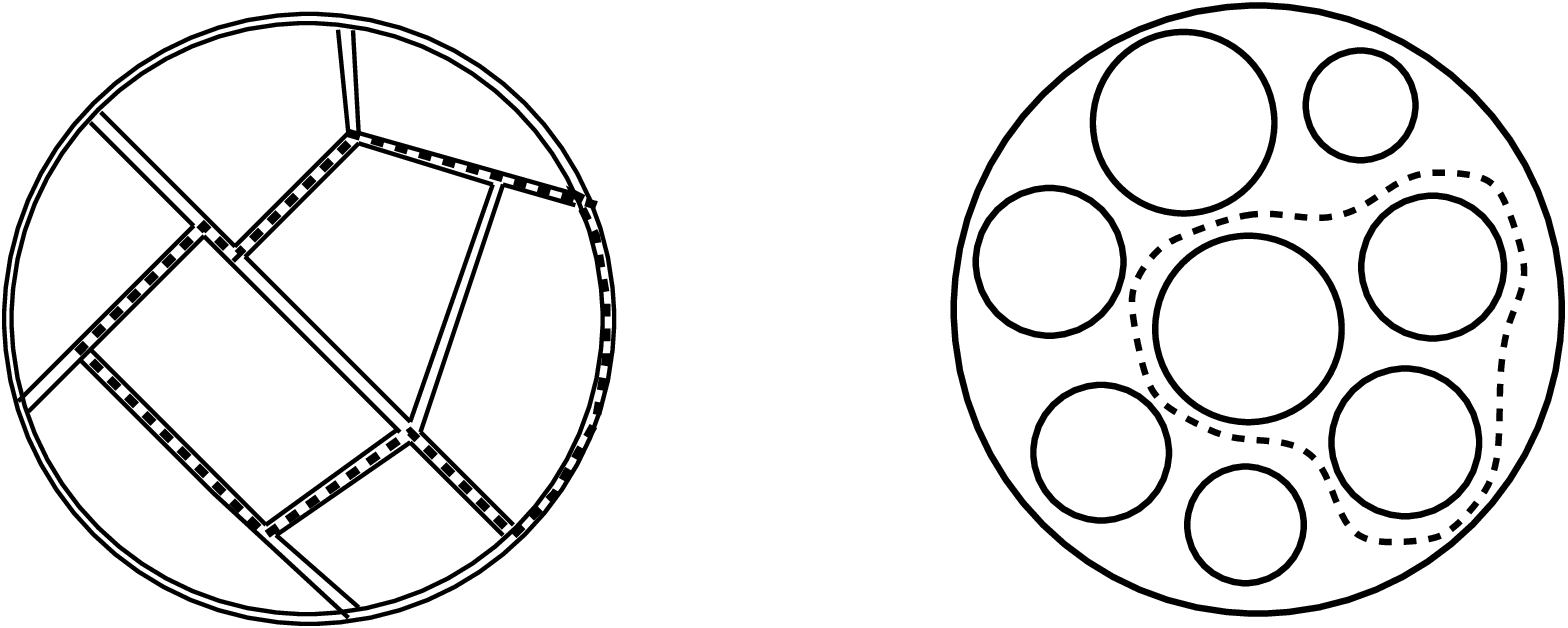}{9cm}{{\bf Fig 3}: In the zero slope limit $\alpha'\to 0$, the open string loop diagram
reduces to an ordinary planar Feynman graph of the low energy gauge theory. The restriction (\ref{rest}) 
on the minimal geodesic length of the non-contractible contour $C$ then translates into a lower bound
on the sum of the Schwinger parameters $t_i$ of the propagators contained in $C$. }

\noindent
{\bf 3. Equation of Motion and Bogolyubov's Recursion Formula}

\medskip

Although the above line of argument did not make any use of the AdS/CFT correspondence, it
seems still quite essential that the planar diagrams were in fact string world sheets, so that
one can easily visualize the transition from the open to the closed string channel. In
principle, however, it should be straightforward (though presumably quite tedious) to explicitly
take the $\alpha'\to 0$ limit and translate all steps into purely field theoretic 
language. Here we limit ourselves to just a couple of remarks.

As indicated in fig 3, our regulatization method for the open string diagrams
carries over quite directly to ordinary planar diagrams.  Feynman diagrams of
the low energy large $N$ gauge theory
can also be written as integral over a ``moduli space''
of Schwinger parameters, parametrizing the proper length $t_i$ of the
propagators.  The UV regulator (\ref{rest}) thus amounts to the restriction that
for any closed path $C$ on the graph (as in fig 3)
\be
\label{neps}
\sum_{i \in C} t_i > \epsilon.
\ee
where the sum is over all propagators that make up the contour $C$. 
This restriction indeed renders the 
integral UV finite.  

We can now use a similar reasoning as above to try and extract
the $\epsilon$ dependence, by explicitly differentiating the total integral
over all Schwinger parameters with respect to the UV cut-off (\ref{neps}).
The analog of the formula (\ref{prop}) should now be extracted from analysing
the UV limit of the one-loop gauge theory amplitude in a background large $N$
gauge-field $A$, with couplings $\phi^i$ turned on; 
equation (\ref{prop}) then corresponds to the fact that, to leading 
order in $1/N$, this amplitude factorizes into a sum over gauge invariant 
single trace-operators $O_i$.

Useful insight into how one should interpret the sigma model data contained in
$\Phi$ is obtained by considering the equation of motion of the total effective
action (\ref{gsum}).  It is possible to write it in the form of a recursion
relation, by expanding the closed string background $\Phi$ in powers of the
string coupling $\lambda$
\be
\Phi=\sum_{\rm n\geq1}\,\lambda^{\rm n}\,\Phi_{\rm n}
\ee
where $\Phi_{\rm n}$ is assumed to be independent of $\lambda$. The equation of motion of 
$\Phi_{\rm n}$ 
\be
\frac{\delta S}{\delta\Phi_{\rm n}}\, = \, 0
\ee
now takes the following form
\be
\label{recursion}
\,Q|\Phi_{\rm n} \rangle \, =\, [ \; 1 \; ]_{\rm n} \; + \, \sum_{1\leq m\leq n-1}%
\,\,\sum_{\sum_{\rm j} \,{\rm j} \, k_{\rm j}={\rm m}}\frac{1}{k_{1}!\cdots k_{\rm m}!}\,
[ \, (\Phi_{1})^{k_{1}}\ldots(\Phi_{\rm m})^{k_{\rm m}} ]_{\rm n-m}.
\ee
Here $[ \ldots ]_{\rm n}$ denotes the state associated to a surface as indicated in fig 4:
the sphere with $n$ holes at the end of a tube with length $1/\epsilon$, and with operator
insertions specified by the $(\ldots)$. The above formula can be used to recursively construct
$\Phi_{\rm n}$ from the previous $\Phi_{\rm m}$'s with ${\rm m}<{\rm n}$.

\vspace{-4mm}

\figuur{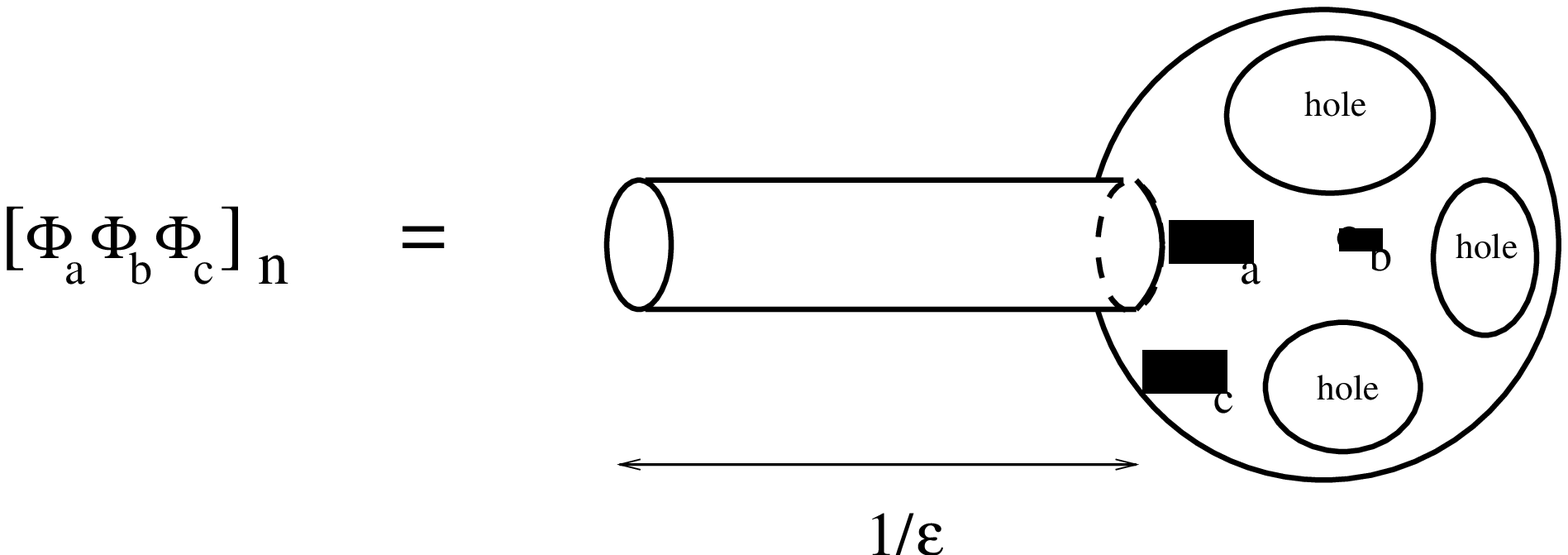}{10cm}{{\bf Fig 4}: This figure illustrates the notation used in (\ref{recursion}):
$[ \, \ldots ]_{\rm n}$ denotes the state associated with the operators
$(\ldots)$ inserted on a sphere with $n$ holes at the end of a tube of length $1/\epsilon$. }

The field theoretic meaning of the above recursive form of the equation of motion
is as follows:  the term $\Phi_{\rm n}$ is associated with the n-th order counter
term designed to cancel the divergent contribution $[\; 1 \; ]_{\rm n}$ of the
n-loop vacuum graph, that is left after subtracting all sub-divergences of lower
order.  There is indeed a strong resemblance between eqn (\ref{recursion}) and
the famous recursion relation due to Bogolyubov \cite{bogol} for constructing the
successive counterterms in the BPHZ renormalization method of QFT.  We suspect
that, by carefully taking the $\alpha' \to 0$ limit of (\ref{recursion}), this
match can be made even more precise.\footnote{It is also not entirely
coincidental that the general solution to Bogolyubov's recursion relation, due to
Zimmerman \cite{zimm}, is known as the {\it forest formula}.  In the present
context, a {\it forest} (a nested or disjoint set of sub-diagrams associated with
nested or disjoint sub-divergences) in essence corresponds to a set of sub-trees
of a closed string tree diagram.  }

\bigskip
\bigskip
\newcommand{\CCC}{{\mbox{\large $\gamma$}}}

\noindent
{\bf 4. Conclusion: World-sheet vs Space-Time RG}

\medskip

Our main result is equation (\ref{final}) and the fact that it reproduces (at
least in form and with the identification $\Gamma_0(\phi) = S_E(\phi)$) the flow
equation (\ref{bflo}) derived from classical 5-d supergravity.
Both derivations have their limited range of validity:  the AdS/CFT duality used
in \cite{vv} is accurate only the regime of $\lambda >\!\!  > 1$, whereas the
perturbative reasoning of this paper requires $\lambda <\!  \!  < 1$.  The
correspondence between the two, however, provides strong evidence that the same
structure persists for all values of the 't Hooft coupling $\lambda$.

Equation (\ref{final}) has an interpretation as both a world-sheet and
space-time RG relation.\footnote{Suggestions as well as concrete proposals
towards a correspondence between world-sheet and space-time RG were made by many
authors, including in \cite{gkp}, \cite{ram}, \cite{p93}, \cite{gomez} and
\cite{christof}.}  The first interpretation is manifest from our derivation,
since the $\beta^i(\phi)$ represent the world-sheet scale dependence of the
sigma-model couplings $\phi^i$.  The second interpretation arises via the
identification of space-time RG transformations with Weyl rescalings of the 4-d
target-space metric $g^{\mu\nu}$.  This motivates the following translation code
between the world-sheet and space-time beta-functions (cf.  \cite{vv}) 
\ba
\label{cflow} 
\beta^i_{{}_{\rm WS}} \is \CCC \, \beta^i_{{}_{\rm ST}} \\[3.5mm]
\beta_{{}_{\rm WS}}^{\mu\nu} \is 2\CCC \, g^{\mu\nu} \!  + \CCC\, 
\beta_{{}_{\rm ST}}^{\mu\nu} 
\label{dflow} 
\ea 
where $\beta_{{}_{WS}}^{\mu\nu}$ denotes the
world-sheet beta-function of the target-space metric $g^{\mu\nu}$. Via
the above translation, the invariance condition (\ref{final}) takes the form 
of a quite conventional RG-flow equation in space-time. 

Note that in this correspondence the world-sheet and space-time scales are
proportional to each other, so UV and UV are mapped to each other as well as IR
to IR.  We further see from (\ref{dflow}) that the prefactor $\CCC$, which
determines the relative normalization of the world-sheet and space-time
beta-functions, is (via eqn (\ref{bg})) proportional to the cosmological term in
$\Gamma_0(\phi)$.  This term, generated by the RR background as well as by other
expectation values, is typically of order the fundamental string scale and can be
even larger.  Although, as seen from \cite{vv}, the interpretation of the 4-d
scale as a holographic extra dimension is expected to break down in this regime,
we have shown here that the RG-invariance (\ref{final}) of $S$ is nonetheless
preserved.

\bigskip
\bigskip

\noindent {\sc Acknowledgements} \\ This work is supported by NSF-grant
98-02484.  We would like to thank M.  Berkooz, R.  Dijkgraaf, G.  Lifschytz, V.
Periwal, A.  Polyakov and E.  Verlinde for useful discussions.  J.K.  is
supported by the Natural Sciences and Engineering Research Council of Canada.

\end{document}